\documentstyle[12pt]{article}
\begin{document}

\author{Guang-jiong Ni \thanks{Email: gjni@fudan.ac.cn} \\
Department of Physics, Fudan University \\ Shanghai
200433, P. R. China}
\title{
To Enjoy the Morning Flower in the Evening - What does the
Appearance of
Infinity in Physics Imply? \thanks{This paper is the English version of its
original article published in the
Chinese journal Kexue ( Science ) Vol. 50, No. 3, 36(1998).
}}
\date{}
\maketitle

\begin{abstract}
A new regularization - renormalization method with no explicit divergence ,
no \hspace{0in}counterterm , no bare parameter and no arbitrary running mass
scale is discussed . There is no difficulty of triviality and the Higgs mass
in the standard model is calculated to be $138Gev$ .
\end{abstract}

\section{ Introduction}

\hspace{0in} \hspace{0in} \hspace{0in} Beginning from the 1930s , the
Quantum Mechanics (QM) evolved into the Quantum Field Theory (QFT) . Not
only the electromagnetic field is quantized into the photon , but also the
electron is further quantized in the second quantization . Both of them
might be created or annihilated in mutual interaction process . Feynman
innovated a diagram method to describe such kind of process , where some
closed loop may occur in the diagram . In the covariant form of QFT , the
particle on the loop is ``virtual''in the sense of the relation among its
mass , energy and momentum in free motion being violated. It is said that
the particle is ``off mass-shell''.The internal momentum of virtual particle
should be integrated in the calculation , then one encountered the problem
of divergence , i.e. , the infinity , in physics .

For example , in the calculation of ``self-energy'' of election in Quantum
ElectroDynamics ( QED) , one considered a freely moving electron with (four
dimensional) momentum $p$. The electron emits a virtual photon with momentum 
$k$ and then absorbs it . This process may induce a modification of electron
mass : $m\rightarrow m$+$\delta m$ , with $\delta m$ being the ``radiative
correction'' of mass . The momentum is conserved at the two vertices . Thus
the Feynman Diagram Itegral (FDI) reads , ($e<0,\hbar =c=1$) :

\begin{equation}
-i\Sigma (p)=-e^2\int_0^1dx\left[ -2\left( 1-x\right) \gamma _\mu p^\mu
+4m\right] I
\end{equation}

\begin{equation}
I\hspace{0in}=\int \frac{d^4K}{\left( 2\pi \right) ^4}\frac 1{\left(
K^2-M^2\right) ^2}
\end{equation}

\hspace{0in}\hspace{0in} 
\begin{equation}
M^2=p^2x^2+\left( m^2-p^2\right) x
\end{equation}

where $x$ is the Feynman parameter , $K=k-xp.$

We shall concentrate on the integral $I$ with the integration range of $K$
being $\left( -\infty \rightarrow \infty \right) $ . Since the denominator $%
\left( K^2-M^2\right) ^2\sim K^4$, so the integral diverges as
\begin{equation}
I\sim \int \frac{dK^4}{K^4}\hspace{0in}\sim \int_0^\Lambda \frac{dK}K
\end{equation}
where a radius $\Lambda $ of large sphere in four dimensional
space is introduced \hspace{0in}and is called as the cutoff of momentum in
integration . When $\Lambda \rightarrow \infty $ , Eq.(4) shows the
logarithmic divergence of most weak type . In other FDIs , one may encounter
the linear divergence $\left( \sim \Lambda \right) $ and the quadratic
divergence $\left( \sim \Lambda ^2\right) $\hspace{0in} etc.

\section{How to deal with the divergence?}

The divergence can not correspond to the observable quantity in physics , so
physicists devised a series of tricks to treat it . Let us look at the
self-energy calculation of election as an example again .

First , one needs to control the divergence , i.e.,to perform the
regularization . The introduction of a cutoff $\Lambda $ into the divergent
integral may be the most naive one , but not satisfied due to its violation
of Lorentz covariance in the theory . In recent 20 years , a most popular
method for regularization is the so-called dimensional regularization scheme
, where the dimension of space-time equals to (4-$\varepsilon $) with 
\hspace{0in}$\varepsilon $ being a continuously varying number approaching
to positive zero . Hence the integral $I$ can be formally expressed
analytically by the Gamma function and the divergence is concentrating into
the pole of Gamma function , i.e., into a term like 1/$\varepsilon .$\hspace{%
0in} \hspace{0in}

Next , for cancelling the divergent term 1/$\varepsilon $ ,one introduces a
counterterm.

As a third step , because only the finite parameter ( e.g. , the mass $m$)
can be observed in the experiment whereas now a divergent counterterm
appears , one has to introduce an another divergent bare parameter ( e.g. ,
the bare mass $m_{0}$). The latter is combined with the counterterm
to cancel the divergence and corresponds to the observed mass m$_R,$which
achieves the renormalization precedure .

An another interesting point is as follows . As the dimension of space-time
has been continued analytically from 4 to $\left( 4-\varepsilon \right) ,$%
the originally dimensionless ( in natural unit system $\hbar =c=1$) charge
square e$^2$ acquires dimension . For making it dimensionless again , one
manages to introduce a factor $\mu ^\varepsilon $ with $\mu $ being an
arbitrary parameter with mass dimension . After performing the
renormalization procedure , the renormalized mass $m_R$ or charge $e_R$ will
very with the running of mass scale $\mu $ . The choice of $\mu $ is
ascribed to the choice of renormalization point , which corresponds to
giving the observed physical parameter ( $m$ or $e$ ) a definite explanation
.

\section{The most difficult aspect in learning is inquiry}

During the teaching and research process for over 40 years , I myself and
many students raised the following questions again and again . How can we
understand the physical meaning of changing the dimension of space-time from
4 to $\left( 4-\varepsilon \right) $? Is the mass generated from its "seed",
the bare mass?
Is the charge e evolving from the bare
charge e$_0$ via the screening effect of vacuum polarization ? Why a new
arbitrary mass scale $\mu $ would emerge abruptly from an originally
definite theory after quantization and renormalization ? Furthermore , when
encountering $\left( 1+B\right) $in calculation , one often replaced it by $%
\left( 1-B\right) ^{-1}$. It is surely reasonable if $B$ is much less than 1
. But now $B$ is approaching infinity , is this reasonable ? Moreover ,
While e$^2$ remains finite , the bare one , e$_0^2,$is infinite , but
sometimes we substitute one by another when doing calculation at higher loop
modification. Is this reasonable ? All these questions bothered me deeply .

From 1991-1994 , based on a lot of literatures [H .Epstein and V . Glaser,
Ann . Inst . Henri Poincare 19 , 211 (1973 ) ; J . Collins , Renormalization
, Cambridge University Press . (1984 ) ; J . Glimm and A . Jaffe ,
Collective Papers , Vol . 2 (1985) ; G . Scharf , Finite Electrodynamics ,
Spring-Verlag, Berlin , 1989 ; J . Dutch , F.Krahe and G . Scharf , Phys.
Lett B 258, 457 (1991 ) ; D .Z.Freedmann , K .Johnson and J . I . Lattore ,
Nucl. Phys . B 371, 353 (1992 ) ] , a PhD candidate , Ji-feng Yang proposed
a new regularization-renormalization method [1] . It was rather simple and
effective , substituting the infinity by some arbitrary constants . It was a
break through , the situation was suddenly enlightened , Further , we have
been studying it in applications and achieving a series of new cognition and
results . Let me try to combine them together in the following for
discussing with our readers .

In August 1997 , when I attended an International Conference at Nanjing
University in memory of Professor C . S. Wu , I read the lecture delivered
by Prof . Wu when she visited the Padova University , Italy , in 1984[2] .I
was greatly inspired . She quoted Calileo's saying that a scientist should
go beyond the ``mere think '' and must raise the ``intelligent questions ''
via experiments . Yes , I had been merely thinking for a long time that
``How can we find a new regularization method to deal with the divergence ?
How can we pick up some thing fixed and finite for explaining the
experimental data after the separation of divergence ?'' Yet I had neither
thought about `` Why the divergence emerges ?'' via the experimental facts ,
nor raised even more acute question that ``If the divergent integral does
not give us infinity, What would it yield ?''

Now I know the answer of the last question . It yields `` Nothing '' first
but `` Arbitrary constants ''afterwards . Because if it does yield something
fixed and finite , it must be wrong . The reason is as follows.

Suppose that after the separation of divergence from some divergent integral 
$I$ , a finite contribution survives and is capable of corresponding to a
tiny but definite mass modification $\delta m$ . Then one can write down $%
m=m_0+\delta m$ . Since $m_0$ is nonobservable , one may perform the
calculation of higher order self-energy FDI for many times and then set $m_0$%
approaching to zero . Hence one might claim that the mass $m$ is generated
via the radiative corrections from self-energy FDI .That would be
``something generated from nothing ''. Actually , mass $m$ \hspace{0in}has a
dimension , say , $m$ equals to 3 grams . It cannot be understood to create
a mass with 3 grams from nothing . Only with the existence of an another
definite mass scale --- the standard weight of 1 gram, could the 3 grams be
understood . This comprises a statement of `` principle of relativity '' in
epistemology . In our point of view , it is nothing but the phase transition
of the environment of particle , ( i.e. , the vacuum ) could provide the
second mass scale and thus renders the generation of particle mass $m$
possible . In this case we have to perform the nonperturbative calculation
with the loop number L in FDI adding up to infinity in QFT [3] . On the
other hand , the self-energy FDI with L finite has nothing to do with the
mass generation .

\section{The subtle use of infinity}

Then I suddenly realized that the emergence of infinity given by a divergent
integral is essentially a warning . It warns us that we had been expecting
too much . We had been sticking to calculate obstinately some thing which is
essentially not calculable .

For a long time I paid no attention to this warning . In studying the $%
\lambda \Phi ^4$model , the basis of standard model in particle physics ,
following many physicists , I also introduced a large but fixed cutoff as
the substitute of infinity . I forgot what my mathematics teacher taught me
46 years ago . At that time , my teacher taught me about ``infinity '' as
follows , ``what is it ? You first raise a large but fixed number N \hspace{%
0in}. Then I say it being larger than N . Next you replace N by another N'(
larger than N ) , I say it being even larger than N'. And so on so forth ,
its limit is infinity '' .Therefore , `` infinity '' is not a very large and
fixed number . Rather ,  it is a symbol .

Yes , the physical meaning of infinity in QFT is nothing but some thing we
can't calculate . It implies a kind of ``infinity '' which goes beyond the
present boundary of our knowledge .

In some sense , the use of infinity in QFT just like that of imaginary
number unit $\sqrt{-1}$ in QM [4] . They have the subtlety with different
tunes rendered with equal skill .

\section{New Regularization - renormalization
Method}

With the previous sections as preliminary , the following trick for handling
the divergence will be quite simple and natural . Let us return back to
Eq.(2) .

Taking the partial derivative of divergent integral I with respect to the
parameter $M^2$with dimension of mass square , we get
\begin{equation}
\frac{\partial I}{\partial M^2}=2\int \frac{d^4K}{\left( 2\pi \right) ^4}%
\frac 1{\left( K^2-M^2\right) ^3}
\end{equation}
As the denominator has a behavior of $K^6$\hspace{0in} , the integral
becomes convergent now :
\begin{equation}
\frac{\partial I}{\partial M^2}=\frac{-i}{\left( 4\pi \right) ^2}\frac 1{M^2}
\end{equation}
For going back to $I$ , we integrate Eq.(6) with respect to $M^2$ , yielding
\begin{equation}
I=\frac{-i}{\left( 4\pi \right) ^2}\left( \ln M^2+C_1\right) =\frac{-i}{%
\left( 4\pi \right) ^2}\ln \frac{M^2}{\mu _1^2}
\end{equation}
Here , according to the theory of indefinite integration , an arbitrary
constant $C_1$appears . We rewrite $C_1=-\ln \mu _1^2$ with $\mu _1$carrying
a mass dimension so that the argument of logarithmic function being
dimensionless.

By means of the `` chain approximation '' , we can derive a renormalized
mass $m_R=m+\delta m$ . When the freely moving particle is on the mass-shell
, i.e. , $p^2=m^2$ , we have $\left( \alpha \equiv e^2/4\pi \right) $:
\begin{equation}
\delta m=\frac{\alpha m}{4\pi }\left( 5-3\ln \frac{m^2}{\mu _1^2}\right) 
\end{equation}
The arbitrary constant $\mu _1$ is fixed as follows.
We want the parameter $m$ in the Lagrangian of original theory being still
explained as the observed mass $m_R$ . As mentioned in previous section ,
due to the disability of perturbative QFT to calculate the mass , the latter
can only be fixed by experiment . Hence , the condition $\delta m=0$ gives
\begin{equation}
\ln \frac{m^2}{\mu _1^2}=5/3,\mu _1=e^{-5/6}m
\end{equation}
The condition $m_R=m$ does not imply that we gain nothing from the
calculation of self - energy FDI . When the motion of particle deviates from
the mass - shell , i.e. ,$p^2\neq m$ , the combination of self - energy
formula with other FDI$_s$ in QED is capable of telling us a lot of
knowledge . For instance, we are able to calculate quickly the (qualitative)
energy shift of $2S_{1/2}$ state upward with respect to $2P_{1/2}$ state in
Hydrogen atom being 997 MHz , so called Lamb shift (with experimental result
1057.8 MHz ) [5] . The latter should be viewed as some mass modification for
an electron in binding state. To be precise , by means of perturbative QFT ,
we can evaluate the mass modification but never the mass generation . This
is a cognition we should keep in mind before we set up to use our method .

Based on this cognition , we replace the divergence by arbitrary constant $%
\mu _1$ , then the latter is fixed by the mass $m$ measured in the
experiment . In some sense , this regularization trick renders the
renormalization very easy , only one step is needed before putting into the
place . No ambiguous concept like the counterterm and/or bare parameter is
needed any more .

Here , two remarks are important :

(a) . There was the trick of taking derivative for reducing the degree of
divergence of some integral $I$ in previous literatures (as mentioned in
sect. 3 ) . Only the elementary calculus for a freshman is needed in such
kind of trick , it is rather simple. However, the crucial point lies in
the fact that we should act from the beginning , act before the counterterm
is introduced, act until the bottom is reached . That is , to take
derivative of integral $I$ with respect to $M^2$ ( or to a parameter $\sigma 
$ added by hand , say $M^2\rightarrow M^2+\sigma $ ) enough times until it
becomes convergent , then perform the same times of integration with
respect to $M^2$( or $\sigma $ then setting $\sigma \rightarrow 0$ again )
for going back to $I$ . Now instead of divergence , we obtain some arbitrary
constant $C_i$ . Note that one divergence is now resolved into some
constants to be fixed . Each $C_i$ has its unique meaning and role which
makes the situation much clearer and well under control [6].

(b) . Some reader may comment that ``The previous methods like the
dimensional regularization scheme can also yield the result in conformity
with the experiment . Though some counterterm and bare parameter are
introduced in the intermediate step , they are not observable and don't
matter much . I think your method being not new. ''In our opinion , when
discussing the basic problem in physics , generally speaking , the simpler
method is the more hopeful one to be correct , whereas the tedious one often
proves to be incorrect as shown by many historical facts . Although the
counterterm and bare parameter disappear at the final , the arbitrary
running mass scale $\mu $ survives .The latter is hard to explain . How can
a theory become unfixed to some extent after quantization and
renormalization while it was totally fixed at the classical level ?

\section{Is the $\lambda \Phi ^4$ model trivial?}

In a text book on QFT , the simplest scalar field model , the $\lambda \Phi
^4$model , is often discussed first . It is defined at the classical level
by the Lagrangian density :
\begin{eqnarray}
{\cal L} & = & \left( \frac{\partial \Phi }{\partial t}\right) ^2-\left( \nabla
\Phi \right) ^2-V\left( \Phi \right) \nonumber \\
V(\Phi ) & = & \frac 12m^2\Phi ^2+\frac 1{4!}\lambda \Phi ^4
\end{eqnarray}
where $m$ is the particle mass excited at the symmetric vacuum $\Phi =0$ and 
$\lambda $ is a dimensionless ( in natural unit system ) coupling constant .

However , the $\lambda \Phi ^4$model with wrong sign in mass term is more
useful in the standard model of particle physics , where $V\left( \Phi
\right) $ is modified into

\begin{equation}
V\left( \Phi \right) =-\frac 12\sigma \Phi ^2+\frac 1{4!}\lambda \Phi ^4
\end{equation}

Hence the ground state vacuum with lowest energy will be shifted from $\Phi
=0$ to $\Phi _1=\left( 6\sigma /\lambda \right) ^{\frac 12}$ . It is called
as the spontaneously symmetry breaking (SSB) . The particle excited at the
symmetry broken vacuum state has a mass $m_\sigma =\left( 2\sigma \right)
^{\frac 12}$ .

Once the model undergoes the quantization , a difficulty arises . It was
found that when the cutoff $\Lambda $ goes to infinity , a singularity would
occur in the solution of renormalization group equation ( RGE ) , which
implies that the renormalized coupling constant $\lambda _R$ approaches to
zero . In other words , there would be no interaction among particles . So
it was thought as a difficulty of so called ``triviality '' and the way out
of the difficulty was setting a large value for $\Lambda $ corresponding to
a high energy , say $10^{15}Gev$ approximately . The $\lambda \Phi ^4$model
should be treated as a low energy effective theory .

Things seem not so fair . Is there a similar problem in QED ? Why people
have seldom talked about its difficulty of triviality ? Maybe the conformity
between QED and experiments ( e.g. , the Lamb shift as mentioned in sect. 5
) is too accurate to be doubted .

We have restudied the $\lambda \Phi ^4$model with SSB according to the
present point of view and found that there is no any triviality difficulty
at all . The crucial point lies in the fact that any model in field theory
defined by the Lagrangian at classical level needs to be redefined once it
is quantized . As a metaphor , with a plane ticket at hand , one has to
reconfirm it by a phone call before his departure from the airport . Here ,
our method to `` reconfirm '' is choosing the two values of arbitrary
constants emerged after the integration of FDI such that the position of
broken vacuum $\Phi _1$and the particle mass $m_\sigma $ excited on it
remain unchanged at perturbative QFT level of any order .

Thus suddenly we saw the light . The difference between the $\lambda \Phi ^4$
model with and without SSB lies in the essence that the former provides two
mass scales , $\Phi _1$and $m_\sigma $ , whereas the latter provides only
one , $m$ . Meanwhile, the invariant meaning of constant $\lambda $ in the
Lagrangian is not the coupling constant , the latter will change after
quantization . For example , at one loop ( $L=1$) calculation , it reads

\begin{equation}
\lambda _R=\lambda \left[ 1+\frac{9\lambda }{\left( 32\pi ^2\right) }\right] 
\end{equation}
The invariant meaning of $\lambda $ is nothing but the ratio of two mass
scales:
\begin{equation}
\lambda =3\left( m_\sigma /\Phi _1\right) ^2
\end{equation}
which remains unchanged irrespective of the order
of $L$ even when $L$ approaches to infinity .

It is more interesting to see that when performing the nonperturbative
calculation in QFT , corresponding to add up approximately \hspace{0in}the
contributions of FDI with $L$ approaching to infinity ( e.g. , by RGE method
) , we find a singularity $\mu _c$ , a critical value in energy , emerging
in our theory . If the energy $E>\mu _c$ , the original stable vacuum state $%
\Phi _1$would collapse into the state $\Phi =0$ . It could be called as the
"Symmetry restoration" and the original model with SSB becomes
ineffective . Therefore , we share the same opinion that $\lambda \Phi ^4$
model is a low energy effective theory. However, the limitation
in energy is not stemming from a finite cutoff $\Lambda $ introduced by hand
for avoiding the so-called difficulty of triviality . Rather , it is
inherited from the intrinsic property of the model itself . Through out the
calculations, we let the cutoff $\Lambda $ running to infinity while the
renormalized coupling constant $\lambda _R$ remains normal . For example,
in Eq. (12) with $L=1$ , $\lambda _R$ is neither
zero nor infinite .

The fact that there is no singularity in calculation when the
loop number $L$ is finite whereas some singularity emerges when $L$
approaches to infinity has a deep meaning . As a contrast in
mathematics, the geometric series $S_n=1+r+r^2+\cdot \cdot \cdot \cdot \cdot
\cdot +r^n$ is analytic and has no singularity ( except the point at
infinity) . But if $n$ approaches to infinity , $\lim_{n \rightarrow \infty}
S_n=1/\left( 1-r\right) $ does have a singularity ( a pole ) at
$r=1$ .

In our point of view \hspace{0in}, a physical model with some singularity is
normal and nontrivial . On the other hand , a model without any singularity
must be wrong , meaningless or trivial . The famous Liouville's theorem in
the function theory with complex variable claims that if a function has no
any singularity on the whole closed complex variable plane , it must be a
trivial constant . An important nontrivial theory in physics is the general
relativity, its singularity is nothing but the black hole .

Either the divergence in FDI or the singularity in the model of physics
reminds us again and again of the fact that the world is infinite whereas
our knowledge remains finite . Any model can at most provide some local or
unilateral description of nature and the emergence of \hspace{0in}infinity
or singularity just reflects the limit or boundary of our cognition . A
famous Chinese philosopher Zhuangzi ( 369 BC$\sim $286 BC ) said that
``While my life span remains finite , the knowledge will extend beyond any
boundary .'' What he was pondering is just an eternal contradiction in the
process for cognizing the world by humanbeing .

\section{Can the Higgs mass be predicted?}

The discovery of top quark in 1995 was a great triumph of the standard model
in particle physics once again. The coincidence between the
theory and experiment has been improved year after year . One even can claim
that the main purpose of present experiment will concentrate on searching
for the Higgs particle , which is the consequence of coupling between the
SU(2) $\lambda \Phi ^4$model with SSB and the gauge fields .

Can the mass off Higgs particle , $m_H$ , be predicted ? This proves to be a
challenging problem for theorists . For many years , only the upper bound
and /or lower bound on $m_H$ could be estimated ( see , e.g., M. Sher, Phys.
Rep.(1998) 179 , 273 ) . We had also attacked this problem in 10  years ago
and obtained the following result [7] :
\begin{equation}
76Gev<m_H<170Gev
\end{equation}
where the
lower bound seemed surprisingly high at that time. So some authors
did not believe in it . However , the lower bound given by experiment
already exceeded $65Gev$ in several years ago and now is approaching $90Gev$
. We are unsatisfied with our previous result , Eq(14) , now . In Ref [7]
,we fixed the cutoff $\Lambda $ to a large value and calculated $m_H$ by old
method , which proved to be a strenuous work and not a beautiful one . Now
the situation is totally different . Based on the new point of view and new
method, we arrive at a predicted value [8 , 9]:
\begin{equation}
m_H=138Gev
\end{equation}
quite fluently and clearly .

This value is found with other available experimental data as imput and is
located within the range constrained by the present phenomenological
analysis on experiments .

We believe that the precision of prediction on $m_H$ could be
improved in accompanying with the progress in accuracy of the relevant
experiments . As the energy now accessible by the accelerators is not far
from $138Gev$ , it would be not too far to reveal the mystery of
nature --- if the Higgs particle can really be found in experiment ?

\section{Concluding remarks}

After diving and floating in the QFT for over 40 years , I ``swam''
eventually to a place where I can relax for a while \hspace{0in}. Eventually
I am getting rid of the four puzzles ---the explicit divergence , the
counterterm , the bare parameter and the arbitrary running mass scale ---
which had bothered me for so many years. I begin to understand that ``only
one step is needed before putting into the place '' for renormalization ,
which is actually nothing but a procedure to ``reconfirm '' something . In
the long river for humanbeing to cognize the world , `` the way seems
extremely long '', we shall continue `` to search for (the truth) while
keeping up and down '' (quoted from the poem by the famous Chinese poet and
stateman Qu Yuan , 340 BC---278 BC ) . But `` where we are located roughly
'' seems comparatively clear . I am delighted in my mood which leads to a
poem :

I'm now enjoying ``Four No '' ,

It leaves only one step to go .

How subtle the ``renormalization '' is ,

Just like to ``reconfirm '' or so.

\noindent $\left[ {\it Note}\right] $ ``Four No'' means ``no explicit divergence, no
counterterm, no bare parameter and no arbitrary running mass scale'',
Actually, further ``no triviality'' is also claimed.

\end{document}